%% file: main.tex
\documentclass[10pt,conference]{IEEEtran}

\usepackage{cite}
\usepackage{amsmath,amssymb,amsfonts}
\usepackage{algorithmic}
\usepackage{graphicx}
\usepackage{textcomp}
\usepackage[table]{xcolor}
\usepackage[hyphens]{url}
\usepackage{fancyhdr}
\usepackage{booktabs}
\usepackage{supertabular}
\usepackage{makecell}
\usepackage{enumitem}
\usepackage{listings}
\usepackage{lstlinebgrd}
\usepackage{hyperref}

\newcommand{\projname}{GemAtlas}

\definecolor{ctrlcolor}{rgb}{0.4, 0.4, 0.4}
\definecolor{instcolor}{rgb}{0.0, 0.0, 0.7}
\definecolor{JungleGreen}{rgb}{0.16, 0.67, 0.53}

\definecolor{sassLineBg}{HTML}{EEF5F5}
\definecolor{sassLineNo}{HTML}{777777}
\definecolor{sassOrange}{HTML}{D97800}
\definecolor{sassPurple}{HTML}{5338D6}
\definecolor{sassComment}{HTML}{777777}

\lstdefinelanguage{SASS}{
  sensitive=true,
  morecomment=[l]{//},
  morekeywords=[1]{IMAD,IADD3,S2R,HADD2,MOV,NOP},
  keywordstyle=[1]\color{sassOrange}\bfseries,
  morekeywords=[2]{LDGSTS,LDG,LDG.E,LDC,STG,LDS,ULDC},
  keywordstyle=[2]\color{sassPurple}\bfseries,
}

\lstdefinestyle{sasspretty}{
  language=SASS,
  basicstyle=\ttfamily\fontsize{6.7}{7.0}\selectfont,
  columns=fixed,
  keepspaces=true,
  showstringspaces=false,
  breaklines=true,
  breakatwhitespace=false,
  breakautoindent=true,
  breakindent=2.2em,
  numbers=left,
  numberstyle=\fontsize{5.2}{5.9}\selectfont\color{sassLineNo},
  numbersep=0.65em,
  frame=none,
  captionpos=t,
  abovecaptionskip=0.45em,
  belowcaptionskip=0.2em,
  xleftmargin=1.8em,
  escapeinside={(*}{*)},
  alsoletter={._0123456789},
  linebackgroundcolor={\color{sassLineBg}},
  linebackgroundsep=0.12em,
  linebackgroundwidth=\dimexpr\linewidth+0.12em\relax,
  linebackgroundheight=1.05\baselineskip,
  linebackgrounddepth=0.25\baselineskip,
  commentstyle=\color{sassComment}\itshape,
}
\newsavebox{\lstscalebox}
\newcommand{\lstscale}{0.83}

\makeatletter
\renewcommand*\lst@makecaption[2]{%
  \vskip\abovecaptionskip
  {\normalfont\fontsize{9.64}{11.25}\selectfont\centering #1. #2\par}%
  \vskip\belowcaptionskip
}
\makeatother

\title{Uncovering the Memory Hierarchy of Modern NVIDIA GPUs for Accurate Simulation}

\author{
\IEEEauthorblockN{Ziwen Wang\textsuperscript{1}, Jiapeng Zhou\textsuperscript{1}, Changxi Liu\textsuperscript{1,*}, Dongsheng Wang\textsuperscript{2}, Yun Chen\textsuperscript{1,*}}
\IEEEauthorblockA{\textsuperscript{1}The Hong Kong University of Science and Technology (Guangzhou), \textsuperscript{2}Tsinghua University\\
\{zwang578, jzhou418\}@connect.hkust-gz.edu.cn, \{changxiliu, yunchen\}@hkust-gz.edu.cn, wds@tsinghua.edu.cn\\
\textsuperscript{*}Corresponding authors}
}

\begin{document}
\maketitle
\thispagestyle{plain}
\pagestyle{plain}

\begin{abstract}
Emerging artificial intelligence workloads, particularly Large Language Models (LLMs), place unprecedented demands on GPU memory systems. Consequently, architectural innovations increasingly target the memory system to alleviate these bottlenecks. However, existing works still rely on outdated memory hierarchy models that fail to capture the complexities of modern hardware. This fidelity gap leads to inaccurate characterization and could misguide optimization strategies.

To bridge this gap, we present \projname{}, the first systematic reverse engineering of the memory hierarchy in modern NVIDIA GPUs.
By developing fine-grained SASS-level benchmarking, we expose critical, previously undocumented micro-architectural details across different cache levels, including cache set associativity, slice-routing and set-indexing hash functions, replacement policies, coherence behavior, and interconnect topologies.

We integrate these discovered micro-architecture details into the state-of-the-art GPU simulation framework.
The final results demonstrate that \projname{} reduces the mean absolute percentage error (MAPE) from {31.69\%} to {22.24\%} {on average across four GPUs}. Specifically, \projname{} reduces the MAPE from {34.04\%} to {23.98\%} for real-world LLM operators.%
\end{abstract}

\input{sections/introduction}
\input{sections/background}

\input{sections/motivation}
\input{sections/reverse_engineering_methodoloy}
\input{sections/L1Cache}

\input{sections/L2}

\input{sections/CCache}
\input{sections/modeling}
\input{sections/discussion}

\input{sections/relatedwork}
\input{sections/conclusion}

\bibliographystyle{IEEEtranS}
\bibliography{refs}

\end{document}

%% file: sections/introduction.tex
\section{Introduction}
\label{sec:introduction}

In recent years, GPUs have evolved from graphics accelerators into general-purpose computing platforms and are now widely used for throughput-oriented workloads, including bioinformatics~\cite{huerta_ref_24,huerta_ref_65}, molecular dynamics~\cite{huerta_ref_56,huerta_ref_98}, reacting-flow simulation~\cite{huerta_ref_41}, and deep learning~\cite{huerta_ref_28,huerta_ref_64,achiam2023gpt}. As GPU computing capabilities continue to advance, improvements in the memory system have not kept pace, making memory performance an increasingly important bottleneck\cite{gholami2024ai}.

More recently, LLM training and inference have become the primary workloads on modern GPUs. As model sizes and context lengths increase, the memory footprint of model parameters and the KV cache imposes high bandwidth and capacity demands on the memory subsystem~\cite{chowdhury2026io, li2024serving}. 
Consequently, overall performance of GPU system has become increasingly sensitive to memory hierarchy behavior and data path efficiency.

However, the memory hierarchy of modern commercial GPUs remains largely opaque at the microarchitectural level. Existing reverse-engineering studies on modern GPUs have focused on SM-core internals, providing detailed analyses of scheduling and execution behavior~\cite{huerta2025}. In contrast, reverse engineering of the memory hierarchy remains fragmented and coarse-grained~\cite{prior_work_2,huerta_ref_58,huerta_ref_97,huerta_ref_54,noc_paper,MT4G}, making it difficult to explain memory-dependent performance behavior on real hardware and perform fine-grained performance optimization. %

To address this gap, we present \projname{}, a unified reverse-engineering and modeling framework dedicated to the memory hierarchy of modern GPUs. \projname{} characterizes key memory-hierarchy behaviors at cache-set granularity across the L1, L2 data cache, and constant-memory hierarchies (L0, L1, and L1.5).
Table~\ref{tab:wide_table} summarizes our specific reverse-engineering contributions across these levels, detailing newly exposed set associativity, topologies, mapping policies, and replacement mechanisms. To validate our findings and evaluate their impact, we incorporate them into AccelSim, a widely used GPU simulator~\cite{accel_sim,huerta2025}.
Evaluation on a benchmark suite spanning traditional GPU benchmarks and real-world LLM operators shows substantially improved fidelity, reducing the average MAPE {over four GPUs} from {31.69\%} to {22.24\%}, and from {34.04\%} to {23.98\%} for the LLM operators, with particularly large gains for memory-sensitive workloads, whose MAPE decreases from {36.33\%} to {20.61\%}. In summary, \projname{} makes the following contributions:
\definecolor{mygreen}{RGB}{0,176,80}
\newcommand{\First}{\textcolor{mygreen}{\textbf{\checkmark$^{\mathbf{N}}$}}} 
\newcommand{\Improve}{\textcolor{orange!90!black}{\textbf{\checkmark$^{\mathbf{I}}$}}} 
\newcommand{\NoWork}{\textcolor{black!35}{\textbf{$\times$}}} 
\newcommand{\NAcell}{\textcolor{black!20}{\textbf{--}}}       

\newcommand{\GroupHeader}[1]{\rowcolor{black!8}[0pt][0pt]\multicolumn{6}{@{}l@{}}{\textbf{#1}}\\}

\begin{table*}[t]
\centering
\caption{Contribution map of reverse-engineered mechanisms on H20 and H100 across cache levels.}
\label{tab:wide_table}
\footnotesize
\setlength{\tabcolsep}{7.5pt}
\renewcommand{\arraystretch}{1.05}

\begin{tabular}{@{} l l ccccc @{}}
\toprule
\multicolumn{2}{@{}l}{\textbf{Reverse-Engineered Mechanism}} & \textbf{L1 D-Cache/SMEM} & \textbf{L2 Cache} & \textbf{L0 C-Cache} & \textbf{L1 C-Cache} & \textbf{GCC (L1.5)} \\
\midrule
\textbf{Structure} & Hierarchy Topology      & \First & \Improve & \NAcell   & \NAcell   & \NoWork \\
 & Capacity                                 & \Improve & \NoWork & \First   & \Improve   & \Improve \\
 & Geometry (line/set/way)                  & \Improve & \Improve & \First   & \Improve & \Improve \\
\midrule
\textbf{Mapping \&} & Physical Routing (slice) Hash Function                    & \First   & \First   & \NAcell   & \NAcell   & \NoWork   \\
\textbf{Policy} & Cache Set Index Hash Function                   & \First   & \NoWork   & \First   & \Improve & \First   \\
 & Cache Set Replacement Policy  & \First   & \Improve & \First   & \First   & \NoWork \\
 & Cache-Set-Level Cross-Partition Coherence     & \NAcell  & \First   & \NAcell  & \NAcell  & \NAcell  \\
\midrule
\textbf{Validation \&} & Latency \& Data-Path Characterization  & \Improve & \Improve & \Improve & \Improve & \Improve \\
\textbf{Utility} & Eviction-set Construction Support        & \First   & \First   & \First   & \First   & \NoWork  \\
\bottomrule
\addlinespace[0.6ex]
\multicolumn{7}{@{}l@{}}{%
  \scriptsize \textbf{Legend:} \quad
  \First~New (first reverse-engineered) \qquad
  \Improve~Improved over prior work \qquad
  \NoWork~Not covered \qquad
  \NAcell~Not applicable%
}
\end{tabular}
\end{table*}

\begin{itemize}[leftmargin=*,nosep]
    \item We reverse-engineer the L1 data cache/shared memory microarchitecture on {H20, H100, A800, and RTX 4090} and disclose its sliced organization, slice routing and cache set indexing, tag--data asymmetry, and replacement policy.
    \item We reconstruct the complete topology of the distributed Hopper L2 cache across partition, Frame Buffer Partitions (FBP) and slice levels; derive the memory-interleave hash; expose coherence/replacement behavior; and introduce an efficient hash-guided eviction-set construction method for modern GPUs.
    \item We systematically dissect the constant memory hierarchy across L0, L1, and GPC-level L1.5 cache, recovering hierarchy organization, cache-set hash indexing rules, and replacement policy.
    \item We develop a benchmark suite, \projname{}-Bench, that extends GPU simulation beyond traditional workloads to include real-world LLM operators drawn from vLLM~\cite{kwon2023efficient} and PyTorch~\cite{paszke2019pytorch}.
    \item We integrate the recovered mechanisms into AccelSim~\cite{accel_sim}, a widely used GPU simulator,
    and validate the model against {four GPUs (H20, H100, A800, and RTX 4090)} using \projname{}-Bench,
    substantially improving accuracy for LLM operators and memory-sensitive workloads.
\end{itemize}

%% file: sections/background.tex
\section{Background and Motivation}
\label{sec:background}

\textbf{GPU Architecture and Memory Hierarchy.}
Modern NVIDIA GPUs are organized hierarchically into Graphics Processing Clusters (GPCs), Streaming Multiprocessors (SMs), and SM-subpartitions (SMSPs)~\cite{huerta_ref_55,H100_white_paper}. In the CUDA programming model, a kernel is decomposed into thread blocks, blocks are scheduled onto SMs, and warps are the basic scheduling unit within an SMSP. The on-chip memory hierarchy consists of three parts. Each SM has a unified L1 cache that integrates data caching, shared memory (SMEM), and texture functionality~\cite{H100_white_paper}. A dedicated constant-memory hierarchy provides access paths for read-only data through the Indexed Constant Cache (IDC), Data Constant Cache (DCC), Immediate Constant Cache (IMC), and GPC Constant Cache (GCC)~\cite{ncu_profiling_guide}. All SMs access a shared L2 cache through the on-chip network (NoC). This L2 is distributed across partitions, each connected to a subset of GPCs~\cite{noc_paper,liu2026dgna}. Each partition contains multiple Frame Buffer Partitions (FBPs) that interface with the memory controllers directly, and each FBP groups multiple independently addressed L2 slices.

\textbf{Undocumented Memory Hierarchy Details.} Modern GPU workloads, particularly LLMs, are increasingly bottlenecked by memory hierarchy behavior~\cite{gholami2024ai}, and architecture-specific optimizations that explicitly exploit it have delivered substantial performance gains~\cite{fu2023autoscratch,dong2026accelerating}. {Newer GPU features, such as thread block clusters with multicast, L2 cache residency control, and the Tensor Memory Accelerator (TMA), also build on this hierarchy. Understanding the memory hierarchy is therefore a prerequisite both for optimizing these workloads and for studying and modeling these features.} Yet the memory hierarchy remains opaque. Official documentation describes only coarse-grained properties, such as cache capacities~\cite{H100_white_paper}. Prior reverse-engineering studies go further and measure finer-grained properties, such as line sizes, latencies, and the associativity of some caches~\cite{huerta_ref_97,huerta_ref_47,huerta_ref_46,huerta_ref_48,prior_work_2,noc_paper,huerta_ref_58,MT4G}. However, for the large, sliced caches of recent GPUs, set-level mechanisms, such as address mapping and replacement policies, remain unknown. As a result, \textbf{developers cannot reliably predict the performance impact of high-level optimizations, making performance tuning inherently empirical.}

%% file: sections/motivation.tex
\label{sec:motivation}

In addition, the lack of microarchitectural understanding limits architecture research. Existing GPU simulators~\cite{gpgpu_sim,accel_sim,huerta2025} rely on simplified or outdated memory-hierarchy models and have been evaluated primarily on traditional benchmark suites. Although some evaluations include GEMM kernels, modern LLM workloads comprise a much broader range of operators with diverse memory behavior, leaving their performance simulation unexplored. Our evaluation shows that existing simulators are inaccurate on these operators:
\begin{table}[h]
\centering
\normalsize
\begin{tabular}{p{0.95\linewidth}}
\textit{On the real-world LLM operators,
the newest Accel-Sim GPU simulator~\cite{huerta2025,accel_sim} has 39.91\% MAPE, with MAPE reaching {56.66\%} for {the {11}} memory-sensitive {LLM} operators}.
\end{tabular}
\end{table}

\noindent{}Consequently, the fidelity of existing simulators on modern GPU workloads remains uncertain, undermining the reliability of architecture exploration.

Together, these limitations motivate systematic reverse engineering of the modern GPU memory hierarchy and its accurate modeling in GPU simulators. By exposing the underlying mechanisms, we aim to bridge this knowledge gap and establish a high-fidelity foundation for systematic performance optimization and architectural analysis.

%% file: sections/reverse_engineering_methodoloy.tex
\section{Reverse Engineering Methodology}
\label{sec:reverse_engineering_methodology}

{
\subsection{The \projname{} Framework}
\label{subsec:framework}
\projname{} is a fully automated framework that performs reverse engineering end to end and feeds the recovered parameters directly into simulation. It consists of three components: (1) a SASS microbenchmark suite that implements every probe in Sections~\ref{sec:l1_microarchitecture}--\ref{sec:constant_hierarchy} as a parameterized template and assembles it for the target GPU; (2) an analysis pipeline that runs the suite and applies the inference steps of Sections~\ref{sec:l1_microarchitecture}--\ref{sec:constant_hierarchy} to produce a parameter table for each cache level; and (3) an extended AccelSim whose memory hierarchy is configured directly from this table (Section~\ref{sec:modeling_eval}). For a previously unseen GPU, the user only specifies the target architecture, and probing, inference, and simulator configuration then run without manual intervention.
\par}

\textbf{Test Platform.} We evaluate \projname{} on NVIDIA H20 and H100 GPUs, both based on the Hopper microarchitecture, as well as A800 (Ampere) and RTX 4090 (Ada), using CUDA 12.8 and \texttt{cuAssembler}~\cite{cuassembler}. We use H20 as the primary platform for presenting results and report the results on the other GPUs in Section~\ref{subsec:other_gpus_re}. \projname{} is hardware-agnostic and can be readily extended to Blackwell and future architectures with support for their corresponding SASS ISAs.

\subsection{SASS-Level Microbenchmarking}
\label{subsec:sass_methodology}
CUDA provides a high-level programming abstraction for NVIDIA GPUs, where applications are written in a C-like language and compiled into PTX, a virtual ISA, before being translated into SASS, the native ISA executed by the hardware~\cite{huerta2025}. 
However, neither CUDA nor PTX offers sufficient control for microarchitectural characterization. Compiler optimizations in both the offline backend (\texttt{ptxas}) and the Just-In-Time (JIT) compiler can reorder instructions and alter register allocation, obscuring true latency and invalidating carefully constructed access patterns~\cite{huerta_ref_38}. As a result, deterministic analysis requires direct manipulation of SASS.
Modern NVIDIA GPUs also rely on a compiler-driven \textit{scoreboard} mechanism rather than dynamic hardware scheduling to manage memory dependencies. SASS instructions carry explicit \textit{control codes} that manage dependency barriers and dictate when subsequent instructions may issue~\cite{huerta2025,cuassembler}.

Direct SASS programming enables two key capabilities for microarchitectural characterization. First, explicit control of scoreboarding allows us to serialize dependent operations and measure memory latency with cycle-level precision. Second, fixed instruction streams ensure deterministic code footprints, which are critical for isolating cache behaviors. For example, in the GCC study (Section~\ref{subsec:gcc_cache}), we place instruction lines into a chosen GCC set with a chain of branches to induce instruction-driven cache evictions.
Because NVIDIA does not provide an official SASS assembler, we build our microbenchmarks upon the open-source \texttt{cuAssembler}~\cite{cuassembler}. We extend its Ampere (SM86) support to Hopper (SM90) by updating the ELF parser and packer for the newer \texttt{cubin} format~\cite{huerta_ref_77}, thereby enabling direct SASS execution and microbenchmark construction on Hopper GPUs.

\subsection{Case Study: L1 Eviction Set Construction}
\label{subsec:eviction_set}
\begin{figure}[t]
    \centering
    \includegraphics[width=\linewidth]{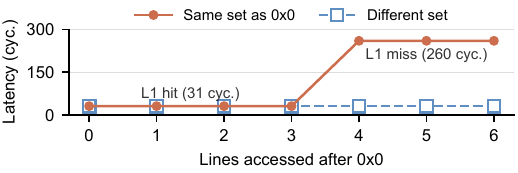}
    \caption{{Latency of re-accessing \texttt{0x0} on H20 after accessing $k$ lines from its eviction set (same set) or from the eviction set of \texttt{0x80} (different set). Each point is the median of 400 runs.}}
    \label{fig:l1_es_latency}
\end{figure}

An eviction set is a group of addresses that map to the same cache set, forming the basis for recovering cache indexing and replacement behavior. As a case study to validate the effectiveness of our SASS-level microbenchmarking framework, we construct L1 eviction sets using the \textit{Group Elimination} algorithm~\cite{cite:group_elimination}, but implemented entirely at the SASS level. The algorithm iteratively prunes a candidate pool while preserving its ability to evict a target line. Each iteration dynamically generates a SASS sequence that accesses the target address $T$ followed by the candidate pool. Using this workflow, we recover stable L1 eviction sets. For example, the eviction set for target address \texttt{0x0} is \{\texttt{0x13880}, \texttt{0x20d80}, \texttt{0x33500}, \texttt{0x4bb00}\}. {Figure~\ref{fig:l1_es_latency} shows the latency of re-accessing \texttt{0x0} as these lines are accessed one by one: \texttt{0x0} remains an L1 hit (31 cycles) after three of them and becomes a miss after the fourth, whereas lines from another set never evict it.} The minimal cardinality of four indicates 4-way associativity and provides the basis for the L1 analyses in the next section.

\begin{lrbox}{\lstscalebox}\begin{minipage}{\linewidth}  %
\begin{lstlisting}[
  style=sasspretty,
  caption={SASS code for an eviction test kernel.},
  label={lst:pchase}
]
[B0-----:R-:W-:-:S04]    NOP;
[B0-----:R-:W0:-:S05]    LDG.E RZ, [R2.64 + T] ;
[B0-----:R-:W0:-:S05]    LDG.E RZ, [R2.64 + CANDIDATE_1] ;
[B0-----:R-:W0:-:S05]    LDG.E RZ, [R2.64 + CANDIDATE_2] ;
[B0-----:R-:W0:-:S05]    LDG.E RZ, [R2.64 + CANDIDATE_3] ;
[B0-----:R-:W0:-:S05]    LDG.E RZ, [R2.64 + CANDIDATE_4] ;
[B0-----:R-:W-:-:S04]    S2R R6, SR_CLOCKLO;
[B------:R-:W0:-:S05]    LDG.E RZ, [R2.64 + T] ;
[B0-----:R-:W1:-:S04]    S2R R7, SR_CLOCKLO;
[B-1----:R-:W2:-:S04]    IADD3 R6, R7, -R6, RZ ;
\end{lstlisting}
\end{minipage}\end{lrbox}
{\centering\scalebox{\lstscale}{\usebox{\lstscalebox}}\par}

Listing~\ref{lst:pchase} shows a representative eviction-test kernel. The kernel operates in three phases: (1) \textit{prime}, loading $T$ into the L1 data cache; (2) \textit{evict}, accessing candidate addresses to create set contention; and (3) \textit{probe}, measuring the latency of re-accessing $T$ using \texttt{SR\_CLOCKLO} and \texttt{IADD3} to classify hits and misses. This execution order is enforced by the scoreboard dependency chain encoded in the bracketed SASS control fields. During traversal, each \texttt{LDG.E} allocates a dependency via \texttt{W0}, while the following instruction waits on it via \texttt{B0}, thereby serializing the loads. The probe path uses the same dependency chain, ensuring that the timing instructions execute only after the re-access to $T$ completes, and their operands are ready. This explicit control yields cycle-accurate hit/miss timing. We use the same SASS-level methodology to construct the eviction sets needed for reverse engineering the L2 data cache and the constant-memory hierarchy in the following sections.

%% file: sections/L1Cache.tex
\section{L1 Data Cache}
\label{sec:l1_microarchitecture}
Modern NVIDIA GPUs use a unified memory structure that dynamically partitions capacity among the L1 data cache, shared memory (SMEM), and texture cache. In this section, we use SASS microbenchmarks to recover the organization of the L1 data cache within this structure. We find that the L1 consists of four slices of 128 4-way sets each, indexed by an XOR hash of the virtual address. It keeps a fixed tag array while its data array shrinks with larger SMEM allocations, and it uses a Segmented LRU (SLRU) replacement policy.

\subsection{Physical Organization and Tag-Data Asymmetry}
\label{subsec:4slice_and_asymmetry}
We begin by characterizing three baseline L1 properties using our SASS-level methodology: line size, associativity, and hit latency. We observe a 128\,B cache line grouped into four 32\,B sectors, 4-way associativity (confirmed by a minimal eviction-set cardinality of four in Section~\ref{subsec:eviction_set}), and a 31-cycle hit latency measured with the SASS probe in Listing~\ref{lst:pchase}.

\begin{table}[htbp]
\centering
\caption{Physical L1 data array capacities under varying shared memory (SMEM) configurations.}
\label{tab:l1_capacity}
\scalebox{0.8}{
\begin{tabular}{@{}lccc@{}}
\toprule
\textbf{Total SMEM} & \textbf{Total L1 Size} & \textbf{Slice Capacity} & \textbf{Data-Line Capacity} \\ \midrule
8\,KiB               & 241\,KiB                & 60.25\,KiB                  & 482                                  \\
16\,KiB              & 233\,KiB                & 58.25\,KiB                  & 466                                  \\
\rowcolor[HTML]{F6F2CB}
32\,KiB     & 217\,KiB (Default)                & 54.25\,KiB                  & 434                                  \\
64\,KiB              & 185\,KiB                & 46.25\,KiB                  & 370                                  \\
100\,KiB             & 149\,KiB                & 37.25\,KiB                  & 298                                  \\
132\,KiB             & 117\,KiB                & 29.25\,KiB                  & 234                                  \\
164\,KiB             & 85\,KiB                 & 21.25\,KiB                  & 170                                  \\
196\,KiB             & 53\,KiB                 & 13.25\,KiB                  & 106                                  \\
228\,KiB             & 21\,KiB                 & 5.25\,KiB                   & 42                                   \\ \bottomrule
\end{tabular}}
\end{table}

\textbf{Accurate Capacity Measurement.}
Prior studies typically estimate L1 capacity using CUDA-level pointer-chasing microbenchmarks~\cite{mei2017dissecting,MT4G}. In contrast, we measure L1 capacity at cache-line granularity using the SASS-level eviction sets derived in Section~\ref{subsec:eviction_set}. For each of the nine SMEM configurations, we fill each cache set until the first capacity-driven eviction and determine the data-array capacity from the number of resident lines. As shown in Table~\ref{tab:l1_capacity}, the combined L1 data cache and SMEM capacity is 249\,KiB in every configuration, 7\,KiB below the 256\,KiB unified capacity reported in official documentation~\cite{H100_white_paper}. Notably, under the default configuration, where the kernel requests no SMEM, the effective L1 data cache capacity is only 217\,KiB, revealing an implicit 32\,KiB reservation for SMEM. We discuss a possible cause of the 7\,KiB gap in Section~\ref{sec:discussion}.

\begin{figure}[htbp]
\centering
\includegraphics[width=0.7\linewidth]{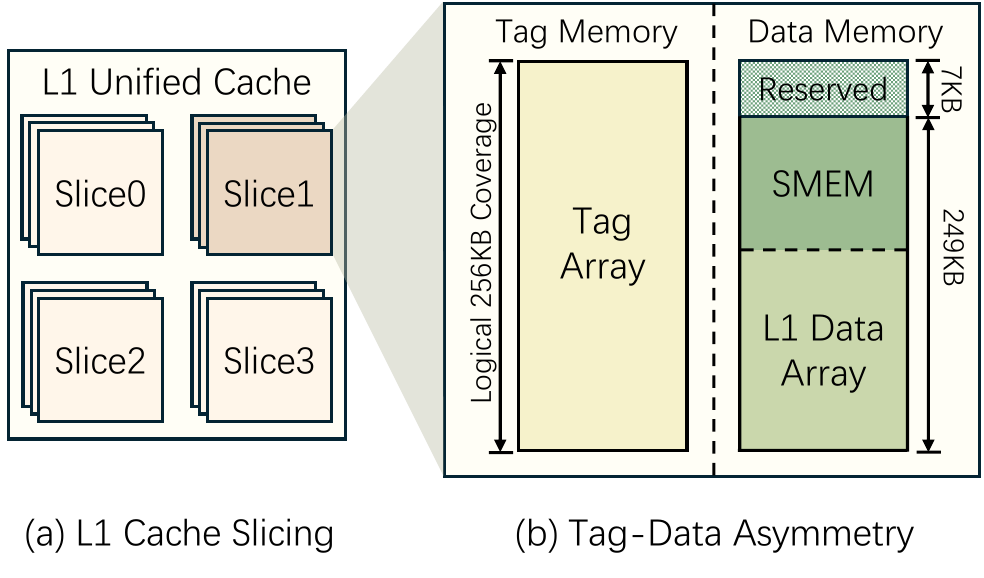}
\caption{Microarchitectural organization of the L1 unified cache.}
\label{fig:L1_Architecture}
\end{figure}

\textbf{The 4-Slice Topology.}
During per-set cache filling, evictions occur before the aggregate L1 capacity is reached. {Moreover, the evicted lines always come from the same one-quarter of the filled sets, for example, 108 of 432 under the default configuration.} %
This indicates that the L1 is divided into four slices, each holding one-quarter of the sets with its own tag and data arrays (Fig.~\ref{fig:L1_Architecture}(a)). Once the data array of one slice is full, a new line in that slice evicts a line from another set of the same slice, while the other slices still have free space. We call this a cross-set eviction and explain it with the tag--data asymmetry (Fig.~\ref{fig:L1_Architecture}(b)) later in this subsection.
Across all SMEM configurations, the four slices have equal capacity, reported in Table~\ref{tab:l1_capacity} in KiB and as the number of 128\,B cache lines per slice.

\textbf{Address Indexing and XOR Hash.}
\label{subsec:hash_routing}
We then recover the L1 indexing logic that maps each virtual address to a slice and an in-slice cache set. Using eviction-set probes at a 128\,B cache-line stride, we find that the 512 cache lines within a 64\,KiB address span map one-to-one onto 512 distinct slice/set pairs, that is, 128 sets in each of the four slices. Each address is therefore mapped by a 9-bit logical index, consisting of a 2-bit slice index and a 7-bit in-slice set index.

\begin{figure}[t]
\centering
\includegraphics[width=\linewidth]{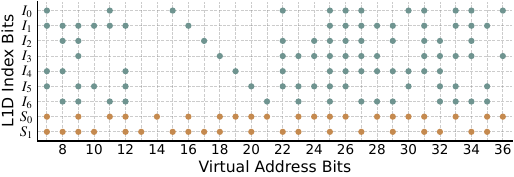}
\caption{Bit-mapping heatmap of the L1 hash-routing function.}
\label{fig:L1_hash}
\end{figure}

Using bit-flipping analysis~\cite{bitflip}, we determine how virtual-address bits generate this 9-bit logical index. The recovered mapping is generated by XOR reductions over selected virtual-address bits drawn from \(b_7\) to \(b_{36}\). Bits \(b_0\)–\(b_6\) do not participate because they encode the 128\,B cache-line offset, and we do not consider bits above \(b_{36}\), which lie outside the 96\,GiB address range of H20.
Figure~\ref{fig:L1_hash} summarizes the recovered bit mapping as a heatmap, where rows $I_0$--$I_6$ correspond to the \textit{in-slice set index} and rows $S_0$ and $S_1$ correspond to the \textit{slice index}. For example, the mapping for cache set index bit $I_0$ is
\[
\begin{aligned}
I_0 ={}&\,b_7 \oplus b_{11} \oplus b_{15} \oplus b_{22} \oplus b_{25} \oplus b_{26} \oplus b_{27}\\
&\oplus b_{30} \oplus b_{31} \oplus b_{33} \oplus b_{34} \oplus b_{36}.
\end{aligned}
\]
{To validate this mapping, we apply it to the measured eviction sets of all 512 sets (15,873 addresses): every address maps to the slice and set of its target. For example, \texttt{0x0} and its eviction set in Figure~\ref{fig:l1_es_latency} all map to slice~0 and set~0.}

\textbf{Hardware Asymmetry: Tag Array vs. Data Array.}
The tag array follows this fixed logical index: its 512 4-way sets with 128\,B lines cover 256\,KiB (Fig.~\ref{fig:L1_Architecture}(b)). We recover identical XOR mappings under all SMEM configurations, so changing the SMEM allocation does not alter this organization. The data array, however, is smaller. Even in the configuration with the largest L1, each slice holds only 482 lines (60.25\,KiB; Table~\ref{tab:l1_capacity}), fewer than the 512 lines needed to fill its 128 4-way sets, and it shrinks further as more of the unified capacity is allocated to SMEM. We refer to this mismatch between a fixed tag array and a smaller data array as the L1 tag--data asymmetry and discuss it further in Section~\ref{sec:discussion}.

This asymmetry leads to three allocation cases. (1) When the target set is full (regardless of whether the data array of the slice is full), the incoming line replaces a line in the same set under the in-set replacement policy (Section~\ref{subsec:cache_replacement}). (2) When the target set has a free tag entry and the slice still has free data capacity, the line is allocated normally. (3) The \textbf{key} case arises when the target set has a free tag entry but the data array of the slice is full, a situation that cannot occur in a conventional cache.
In this case, the hardware performs a \emph{cross-set eviction} within the slice: it first selects a victim set from the sets in the slice, then evicts a line from that set using the local replacement policy, and finally allocates the new line to the target set. {To determine how the victim set is selected, we repeat the per-set filling experiment 13,500 times under the default configuration and record the victim sets in each run. The victim set changes from run to run, and its selection probability depends on the fill order: among the 108 sets in the slice, the first 27 filled are selected about 2.5 times as often as the last 27. All nine SMEM configurations show the same trend. We leave the exact selection rule to future work.}

\subsection{L1 Cache Set Replacement Policy}
\label{subsec:cache_replacement}
\begin{figure}[htb]
    \centering
    \includegraphics[width=\linewidth]{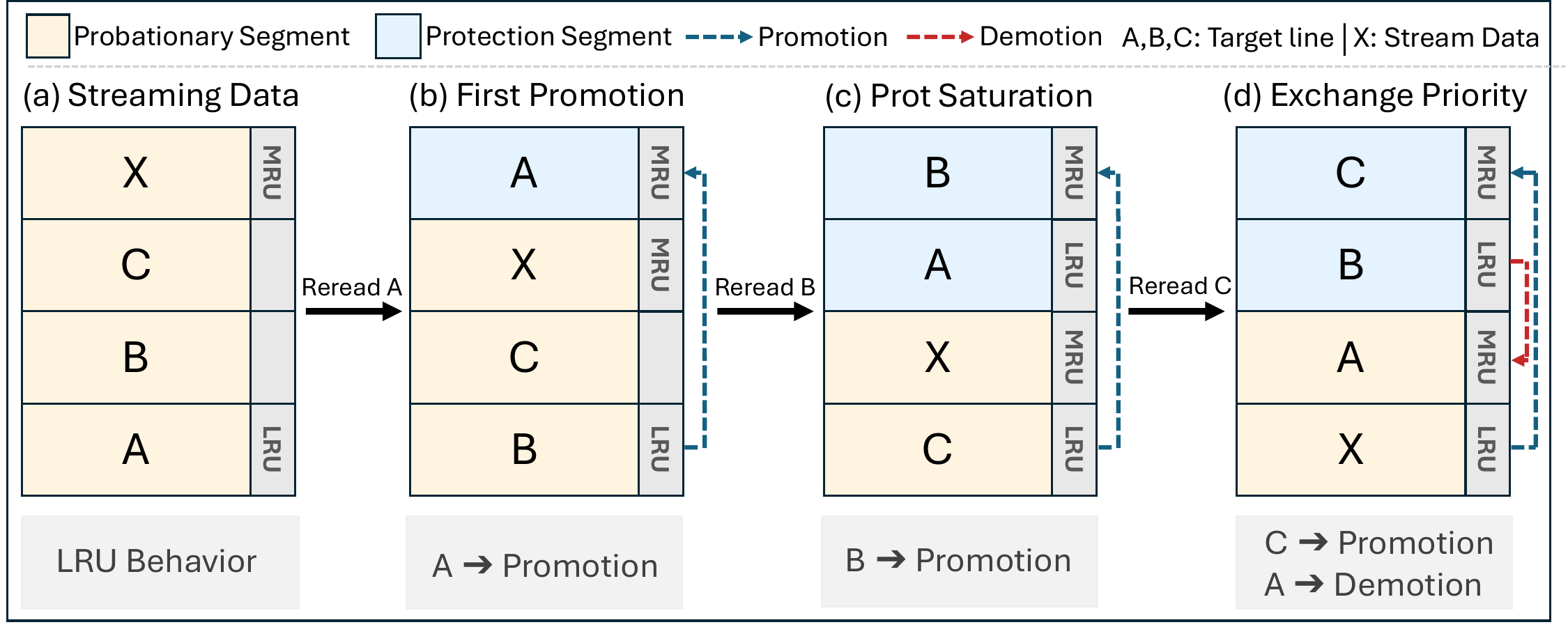}
    \caption{Inferred state transitions of the L1 SLRU policy. Evictions occur exclusively within the Probationary Segment (yellow blocks).}
    \label{fig:L1_SLRU}
\end{figure}

Our measurements show that the L1 data cache employs a Segmented LRU (SLRU) policy~\cite{Gao2010ADS} rather than conventional unified LRU. Each set contains Probationary and Protected segments: new lines enter the Probationary Segment, while re-accesses promote them to the Protected Segment. Figure~\ref{fig:L1_SLRU} summarizes the inferred state transitions.

\begin{lrbox}{\lstscalebox}\begin{minipage}{\linewidth}  %
\begin{lstlisting}[
style=sasspretty,
caption={SASS microbenchmark used to deduce the SLRU mechanisms.},
label={lst:slru_benchmark}
]
// Phase 1: Priming [* = optional re-access]
[B0-----:R-:W0:-:S05]    LDG.E RZ, [R2.64 + ADDR_A];
[B0-----:R-:W0:-:S05]    LDG.E RZ, [R2.64 + ADDR_A]; //*
[B0-----:R-:W0:-:S05]    LDG.E RZ, [R2.64 + ADDR_B];
[B0-----:R-:W0:-:S05]    LDG.E RZ, [R2.64 + ADDR_B]; //*
[B0-----:R-:W0:-:S05]    LDG.E RZ, [R2.64 + ADDR_C];
[B0-----:R-:W0:-:S05]    LDG.E RZ, [R2.64 + ADDR_C]; //*
// Phase 2: Streaming
[B0-----:R-:W0:-:S05]    LDG.E RZ, [R2.64 + ADDR_X];
[B0-----:R-:W0:-:S05]    LDG.E RZ, [R2.64 + ADDR_Y];
[B0-----:R-:W0:-:S05]    LDG.E RZ, [R2.64 + ADDR_Z];
// Phase 3: Probing
[B0-----:R-:W-:-:S04]    S2R R6, SR_CLOCKLO;
[B------:R-:W0:-:S05]    LDG.E RZ, [R2.64 + TARGET_T];
[B0-----:R-:W1:-:S04]    S2R R7, SR_CLOCKLO;
[B-1----:R-:W2:-:S04]    IADD3 R6, R7, -R6, RZ;
\end{lstlisting}
\end{minipage}\end{lrbox}
{\centering\scalebox{\lstscale}{\usebox{\lstscalebox}}\par}

We infer these transitions with the SASS microbenchmark in Listing~\ref{lst:slru_benchmark}, which selectively re-accesses target lines ($A$--$C$), streams new lines ($X$--$Z$) into the same set to trigger eviction, and times a final probe to determine target residency.

\textbf{Streaming-only baseline (Figure~\ref{fig:L1_SLRU}(a)).}
With all optional re-accesses disabled, the access stream is $A,B,C,X,Y,Z$ within one set. We observe that $Y$ and $Z$ evict $A$ and $B$, respectively, consistent with LRU eviction within the Probationary Segment when no line is promoted.

\textbf{Stepwise promotion into Protected (Figure~\ref{fig:L1_SLRU}(b,c)).}
We next introduce re-accesses progressively. When only $A$ is re-accessed (Figure~\ref{fig:L1_SLRU}(b)), $Y$ and $Z$ evict $B$ and $C$ while $A$ remains resident, showing that a re-accessed line is promoted into the Protected Segment and gains higher retention priority. When $B$ is also re-accessed (Figure~\ref{fig:L1_SLRU}(c)), $A$ and $B$ both remain resident while $C$ and $X$ are evicted, indicating that two promoted lines can coexist in the Protected Segment.

\textbf{Priority exchange on protected overflow (Figure~\ref{fig:L1_SLRU}(d)).}
Starting from the state in Figure~\ref{fig:L1_SLRU}(c), we additionally re-access $C$, creating a third promoted line and triggering protected overflow. Under subsequent streaming pressure, we observe that $Y$ evicts $X$ and then $Z$ evicts $A$. This implies that $C$ is promoted into the Protected Segment, the LRU line in the Protected Segment ($A$) is demoted into the Probationary Segment, and the demoted line is inserted at Probationary-MRU, since $A$ survives one insertion longer than $X$. Because the third promotion causes demotion rather than expansion, the Protected Segment can hold at most two lines.

{\textbf{Immediate promotion on re-access.}
We further test whether promotion depends on timing or on access count. (1) Inserting a delay of up to about two million cycles before or after the re-access of $A$ does not change the outcomes in Figure~\ref{fig:L1_SLRU}(b,c), so a line is promoted as soon as it is re-accessed, and its promotion does not expire. (2) Re-accessing $A$ up to four times gives the same outcome as once, so extra accesses do not affect promotion. Moreover, in the overflow test of Figure~\ref{fig:L1_SLRU}(d), $A$ is still demoted when accessed four times and $B$ and $C$ only twice, whereas ordering by access count would demote $B$. Hence, the Protected Segment is ordered by recency, not by access count.}

\textbf{Final replacement policy.}
Taken together, these observations indicate the following SLRU policy: (1) new lines are inserted into the Probationary Segment; (2) a re-access promotes a line into the Protected Segment, where lines have higher retention priority; (3) the two segments maintain separate recency orders, with LRU eviction in the Probationary Segment and, on overflow, the LRU line of the Protected Segment demoted to the MRU position of the Probationary Segment; and (4) the Protected Segment holds at most two lines, and the remaining ways belong to the Probationary Segment.

%% file: sections/L2.tex
\section{Distributed L2 Cache}
\label{sec:l2_reverse}

Modern NVIDIA GPUs (e.g., A100, H100, and B200) organize the L2 cache into two partitions, each connected to a subset of GPCs through the on-chip interconnect~\cite{noc_paper,liu2026dgna}. In this section, we use SASS microbenchmarks to recover this distributed L2 on H20. We find that its 96 slices are grouped into 12 FBPs across the two partitions and are selected by an XOR hash with a 128-to-96 remapping. Using slice-constrained eviction sets built with this hash, we further characterize replacement and set-level cross-partition coherence.

\begin{figure}[ht]
    \centering
    \includegraphics[width=0.85\linewidth]{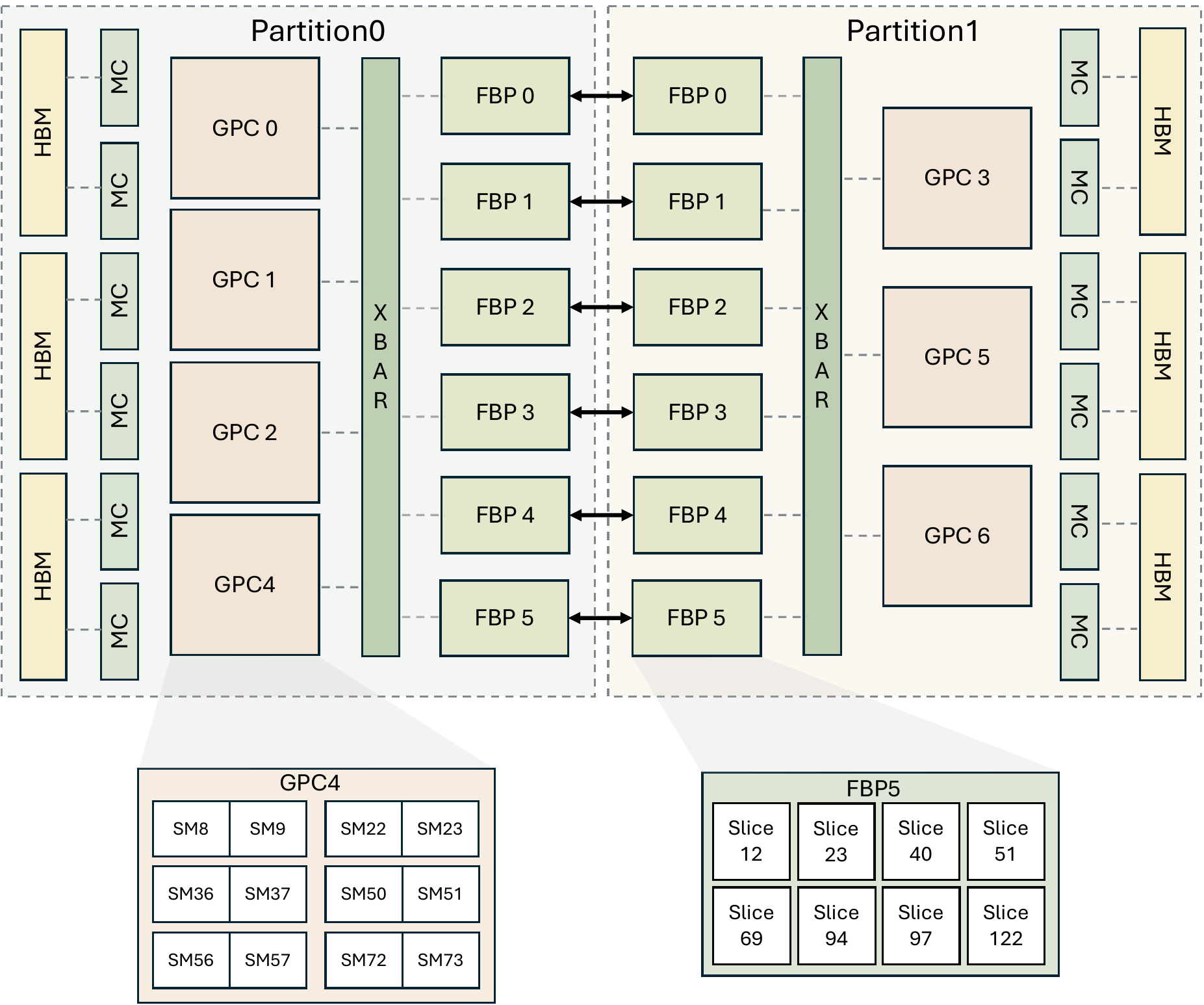}
    \caption{L2 Cache Topology.}
    \label{fig:L2_Topology}
\end{figure}

\subsection{Partition-Level Topology and SM Affinity}
\label{subsec:l2_partition}
\textbf{Mapping SMs to GPCs and Partitions.}
We read both mappings directly from special registers with the SASS \texttt{S2R} instruction. For the SM-to-GPC mapping, \texttt{SR\_AFFINITY} returns the GPC ID of the executing SM. Alternatively, SMs can be grouped by \texttt{SR\_CLOCK}, which, unlike the GPU-wide \texttt{SR\_GLOBALTIMER}, is shared only by the SMs of one GPC (e.g., two GPCs differ by 2,315,338 cycles). Both methods yield the same grouping. For the GPC-to-partition mapping, \texttt{SR\_SW\_SCRATCH} returns the same value for all SMs of a partition: \texttt{0xA0000000} for 42 SMs and \texttt{0x80000000} for the other 36. The resulting topology is asymmetric: Partition0 connects GPC0, GPC1, GPC2, and GPC4, whereas Partition1 connects GPC3, GPC5, and GPC6 (Figure~\ref{fig:L2_Topology}, which also shows the SMs of GPC4).

\textbf{SM-to-Partition Latency.}
Using the SASS probe in Listing~\ref{lst:pchase}, we measure the latency of L2 hits and HBM accesses through the local partition, which connects to the probing SM's GPC, and through the remote one. For L2 hits, we first warm the target line in the designated partition; for HBM accesses, we force an L2 miss. Local and remote L2 hits take $\sim$276 and $\sim$446 cycles, and local and remote HBM accesses take $\sim$530--570 and $\sim$700--750 cycles, respectively. Both therefore incur a similar $\sim$170-cycle cross-partition penalty.

\subsection{L2 Slice and FBP Organization}
\label{subsec:l2_routing}

\textbf{Determining the Number of Physical L2 Slices.}
The number of active physical L2 slices on H20 is determined using a contention-based microbenchmark derived from CuAssembler~\cite{cuassembler}. Global-memory atomic operations execute at the slice where the target cache line is located. Concurrent \texttt{atomicAdd} operations to two addresses in the same slice therefore contend and increase the runtime, whereas addresses in different slices execute nearly in parallel. Runtime thus indicates whether an address pair maps to the same L2 slice.

On H20, we generate 1,024 addresses spanning the range \texttt{0x00000} to \texttt{0x1FF80} (128\,KiB window) with stride \texttt{0x80} (one cache line), and perform pairwise atomic contention tests over all ordered address pairs ($1024\times1024$), treating high-latency pairs as same-slice collisions. Let $N_{\text{collide}}$ denote the number of pairs classified as collisions. {If the addresses are distributed uniformly over $S$ slices, the number of same-slice ordered pairs is
\[
N_{\text{collide}} = 1024^2/S,\quad
S = 1024^2/N_{\text{collide}}.
\]
We measure $N_{\text{collide}} = 12{,}288$, which gives a slice count of $S = 85.333$ on H20.}
This non-integer value is inconsistent with a direct physical indexing scheme. We infer a two-stage remapping design: a 7-bit logical slice space (128 outcomes) followed by remapping onto 96 physical slices. Specifically, 64 slices are uniquely mapped (one logical index each), while the remaining 32 slices are shared by two logical indices. Under this model, the probability of same-slice collisions is \(64\times (1/128)^2 + 32\times (2/128)^2 = 3/256\), which implies an apparent slice count of {\(1/(3/256)=85.333\)}, exactly matching our measurement. We therefore conclude that H20 has 96 active physical L2 slices, with 48 slices per partition and 640\,KiB per slice.

\begin{figure}[h]

    \centering

    \includegraphics[width=0.9\linewidth]{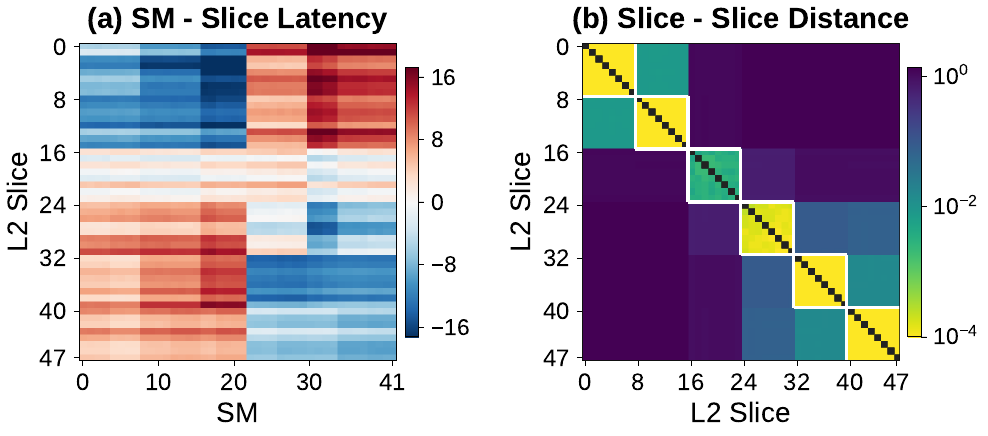}

    \caption{(a) Mean-centered SM-to-slice relative-latency heatmap within one partition. (b) Slice correlation-distance heatmap derived from the latency vectors in (a), where distance is defined as \(1-r\) and \(r\) is the Pearson correlation coefficient. SM labels are display indices reordered according to clustering results (not physical SM IDs).}

    \label{fig:l2_heatmap}

\end{figure}

\textbf{Identifying FBP Groupings via Cross-SM Correlation.}
Slice-access latency alone cannot reliably identify FBP groupings because it also depends on each slice's position within the FBP relative to the XBAR, so slices from different FBPs have overlapping latency ranges~\cite{noc_paper}. Bandwidth contention does not reveal them either, because each memory-side controller sustains more than 300\,GiB/s in our measurements, which our probes cannot saturate. Instead, we identify FBPs from how slice latency varies across SMs, since slices in the same FBP show similar variation. Using the atomic-contention microbenchmark described earlier, we isolate individual slices and measure access latency from each of the 42 SMs in one partition to each of its 48 local slices. After mean-centering per SM to remove systematic offsets, we obtain a \(48 \times 42\) relative-latency matrix (Figure~\ref{fig:l2_heatmap}(a)). We then compute pairwise Pearson correlations between slices and derive the corresponding correlation-distance matrix, shown in Figure~\ref{fig:l2_heatmap}(b). The resulting structure reveals six distinct clusters, each containing eight slices, which is further confirmed by standard clustering analysis. We therefore conclude that each partition consists of 6 FBPs with 8 slices each, yielding a total of 12 FBPs and 96 slices across the GPU.

\textbf{Cross-Partition FBP Correspondence.}
To characterize the cross-partition FBP topology, a fixed SM measures the mean latency to each FBP for remote L2 hits and for the same cache lines after coherence brings them into the home partition. The two sets of latencies have the same rank order across the six FBPs and are highly correlated (Pearson correlation \(>0.95\)), differing by a nearly constant offset of \(\sim\)178 cycles, close to the \(\sim\)170-cycle cross-partition penalty measured in Section~\ref{subsec:l2_partition}. Crossing partitions thus adds an almost uniform penalty without changing the relative SM-to-FBP distances, which implies a one-to-one correspondence between the six FBPs of the two partitions ($P0$ and $P1$). Figure~\ref{fig:L2_Topology} shows this mirror topology as \( \mathrm{FBP}_X(\mathrm{P0}) \leftrightarrow \mathrm{FBP}_X(\mathrm{P1}) \), where the bidirectional arrows denote the cross-partition coherence relationship between matched FBPs.

\subsection{Complete Memory Interleave Hash Recovery}
\label{subsec:l2_hash_recovery}
Using bit-flip analysis~\cite{bitflip}, we recover the complete routing hash that maps each memory address to its partition, FBP, and slice.
Because NVIDIA L2 indexing uses physical addresses while user programs observe only virtual addresses, fully recovering the indexing hash is impossible. We therefore restrict our recovery to a 128\,MiB virtual address window (the low 27 bits), which is sufficient for our subsequent eviction-set construction on the 60\,MiB H20 L2 (Section~\ref{subsec:l2_es}). Modern NVIDIA GPUs commonly allocate memory using 2\,MiB large pages, whose offset bits (bits 0--20) are identical in virtual and physical addresses, so the recovered hash on these bits directly reflects the hardware routing. We therefore report only this portion of the routing hash. The recovered routing equations are:

\begin{equation}
\label{eq:l2_hash_full}
\resizebox{0.88\linewidth}{!}{$ \displaystyle
\begin{aligned}
p   &= PA[12] \oplus PA[13] \oplus PA[15] \oplus PA[17] \oplus PA[19] \\[1ex]
f_0 &= PA[8]  \oplus PA[12] \oplus PA[16] \oplus PA[18] \oplus PA[20] \\
f_1 &= PA[10] \oplus PA[13] \oplus PA[16] \oplus PA[17] \oplus PA[19] \oplus PA[20] \\
f_2 &= PA[11] \oplus PA[14] \oplus PA[18] \oplus PA[19] \\[1ex]
s_0 &= PA[8]  \oplus PA[16] \oplus PA[17] \\
s_1 &= PA[10] \oplus PA[16] \\
s_2 &= PA[11] \oplus PA[16] \oplus PA[17]
\end{aligned}
$}
\end{equation}

These equations define a 7-bit logical routing space consisting of one partition bit \(p\), a 3-bit logical FBP index \(f_{\log}=(f_2f_1f_0)\in\{0,\dots,7\}\), and a 3-bit slice index \(s=(s_2s_1s_0)\in\{0,\dots,7\}\). The FBP and slice equations are shared by both partitions, while \(p\) selects the target partition. Together, they yield \(2\times 8\times 8=128\) logical destinations. However, H20 has only \(2\times 6\times 8=96\) active physical slices (Section~\ref{subsec:l2_routing}), so the 128 logical destinations must be remapped onto them.
{We recover this remap by locating the physical slice of every line in a 2\,MiB page with atomic contention (Section~\ref{subsec:l2_routing}) and comparing it with Eq.~\eqref{eq:l2_hash_full}. The remap preserves the partition bit \(p\). In each partition, six logical FBPs map one-to-one onto the six physical FBPs, and the other two are folded onto two of them with slice index \(s\oplus 4\). For the first 128\,KiB of a page, this gives}
\begin{equation}
\label{eq:remap}
(f_{\mathrm{phys}},\, s_{\mathrm{phys}}) =
\begin{cases}
(f_{\log},\, s), & f_{\log}<6,\\
(f_{\log}\oplus 6,\, s\oplus 4), & f_{\log}\ge 6.
\end{cases}
\end{equation}

{In Eq.~\eqref{eq:remap}, logical FBPs 6 and 7 are folded onto physical FBPs 0 and 1, which therefore receive twice as many lines as the other physical FBPs. In each subsequent 128\,KiB, the folded logical FBPs land on the next pair of physical FBPs, cycling through the three pairs 0--1, 2--3, and 4--5. This modulo-3 rotation balances the load: each physical pair receives the folded logical FBPs in exactly one of every three consecutive 128\,KiB, so all physical FBPs receive the same number of lines over 384\,KiB. Eq.~\eqref{eq:l2_hash_full} and the remap reproduce exactly the 12{,}288 colliding pairs measured in Section~\ref{subsec:l2_routing} ($S = 85.333$).}

\subsection{Slice-Constrained Eviction-Set Construction}
\label{subsec:l2_es}

Prior L2 eviction-set (ES) construction techniques~\cite{cite:group_elimination,invalidate} fail on modern, distributed L2 architectures such as H20. The large capacity of these caches (e.g., 60\,MiB) forces traditional group elimination algorithms to filter very large candidate pools, accumulating noise that discards true ES members. Furthermore, undocumented cross-partition coherence traffic introduces interference into the measurements.

To overcome these challenges, we introduce a slice-constrained group-elimination method. Because only addresses mapping to the same set in the same slice contend, we use our recovered slice hash (Section~\ref{subsec:l2_hash_recovery}) to restrict the candidate pool to the target slice. We also perform profiling from an SM in the same partition, avoiding cross-partition coherence interference. Within this reduced pool, group elimination iteratively discards candidate groups that are unnecessary to evict the target, leaving an ES for the target set. %
{This slice constraint \textbf{aggressively prunes the search space by $93\times$}: in our 120\,MiB candidate pool, only 10,542 of 983,040 lines fall in the target slice.} Successful construction of ESs also provides validation of the recovered slice hash, since stable ESs require a correct slice mapping.

From the resulting ESs, we infer that each H20 L2 cache set is 16-way associative and follows a standard LRU replacement policy. {Specifically, exactly 16 conflicting accesses evict a line, even if it was accessed ten times before, ruling out SLRU. Moreover, when the 16 lines of a set are re-accessed in random orders (900 trials), a new line always evicts the least recently used one, ruling out pseudo-LRU.}

\subsection{Cross-Partition Coherence at Cache-Set Granularity}
\label{subsec:l2_es_coherence}

Using the slice-constrained ESs constructed in Section~\ref{subsec:l2_es}, we first access all lines in an ES from an SM in the home partition, placing them in that partition and verifying that they contend for a single cache set. We then access the same lines from an SM connected to the remote partition to create coherent copies. These copies continue to contend for a single cache set, showing that coherence preserves this cache set indexing rather than distributing the lines across multiple sets. Moreover, all 16 ES lines can be coherently replicated into the remote partition and simultaneously occupy this set. Thus, each cache set can hold any mixture of lines mapped to that partition and coherent copies from the peer partition, with no additional per-set capacity limit beyond the standard 16-way associativity.

Our measurements further indicate that corresponding cache sets in the two partitions maintain independent LRU states. Evicting a line from one set does not change the LRU order or trigger an eviction in the corresponding set of the other partition. During a cross-partition coherence event, the event behaves as a normal cache hit in the home partition and promotes the target line to the most recently used (MRU) position. In the remote partition, the same event behaves as the insertion of a new cache line: the coherent copy is inserted at the MRU position, and if the set is full, its current LRU line is evicted.

%% file: sections/CCache.tex
\section{Constant Memory Hierarchy}
\label{sec:constant_hierarchy}
In modern GPU workloads, constant memory is widely used to store network weights, scaling factors, and kernel parameters. Official documentation~\cite{ncu_profiling_guide} names several constant-cache components (GCC, IDC, DCC, and IMC) but does not describe their organization. In this section, we use SASS microbenchmarks to reconstruct this hierarchy. We find three levels: L0 and L1 constant caches in each SM, both with 2\,KiB, 4-way sets and Tree-PLRU replacement, and a 128\,KiB GCC that is shared by the SMs of each GPC and also holds instructions.

\subsection{The Intra-SM Hierarchy: L0 and L1 Constant Caches}
\label{subsec:intra_sm_constant}
At the SM level, official documentation distinguishes three constant-data access paths, each associated with a different class of SASS instruction~\cite{ncu_profiling_guide}. Immediate operands encoded in instructions (e.g., \texttt{MOV}) use the Immediate Constant Cache (IMC) path, \texttt{ULDC} uses the Data Constant Cache (DCC) path, and standard \texttt{LDC} uses the Indexed Constant Cache (IDC) path. We measure their cache hit latencies as 2, 15, and 25 cycles, respectively.

\textbf{DCC and IMC Share a Single Physical Cache.}
Cross-instruction probing shows that \texttt{MOV} and \texttt{ULDC} share a single cache: warming a constant address with \texttt{MOV} makes a subsequent \texttt{ULDC} access hit, and capacity evictions through \texttt{MOV} also evict data brought in by \texttt{ULDC}, whereas cross-probing between \texttt{LDC} and either \texttt{MOV} or \texttt{ULDC} misses in both directions. Thus, the IMC and DCC are physically unified, while the IDC is separate. We attribute the latency difference between \texttt{MOV} and \texttt{ULDC} to their distinct data paths: \texttt{ULDC} routes data through the uniform-register path, whereas \texttt{MOV} uses direct operand fetch.

\textbf{Cache Placement and Structural Parameters.}
We identify the physical placement of the IDC through SASS-level cross-warp probing: when one warp warms a cache line using \texttt{LDC}, other warps on different sub-partitions within the same SM still observe a hit, indicating that the IDC is shared at the SM level. We therefore classify it as the L1 Constant Cache. In contrast, the physical placement of the IMC/DCC cache cannot be directly determined due to limitations in the current SASS assembler tool. The documentation is also inconsistent: it places the IMC at the SM-subpartition level and the DCC at the SM level, even though our measurements show that the two are physically unified. Given the 2-cycle hit latency of the IMC, we conjecture that this shared cache resides at the SM-subpartition level and treat it as the L0 Constant Cache.

Using SASS-level microbenchmarks, we characterize the structural parameters of the two intra-SM constant caches. The L0 and L1 constant caches exhibit identical structural parameters: each has a 2\,KiB capacity, a 64\,B line size, 4-way associativity, and 8 sets. In both caches, set indexing follows the same linear function, $\text{set\_index} = (\text{offset} / 64) \pmod 8$.

\begin{lrbox}{\lstscalebox}\begin{minipage}{\linewidth}  %
\begin{lstlisting}[
style=sasspretty,
caption={Microbenchmark for reverse-engineering the replacement policy.},
label={lst:tree_plru}
]
// Phase 1: Sequential fill of the 4-way set
[B0-----:R-:W0:-:S05]    LDC R8, c[0x3][ADDR_A] ;
[B0-----:R-:W0:-:S05]    LDC R8, c[0x3][ADDR_B] ;
[B0-----:R-:W0:-:S05]    LDC R8, c[0x3][ADDR_C] ;
[B0-----:R-:W0:-:S05]    LDC R8, c[0x3][ADDR_D] ;
// Phase 2: Selectively re-access
[B0-----:R-:W0:-:S05]    LDC R8, c[0x3][ADDR_X] ;
// Phase 3: Trigger eviction and measure survival
[B0-----:R-:W0:-:S05]    LDC R8, c[0x3][ADDR_E] ;
\end{lstlisting}
\end{minipage}\end{lrbox}
{\centering\scalebox{\lstscale}{\usebox{\lstscalebox}}\par}

\textbf{Tree-PLRU Replacement Policy.}
To determine the cache set replacement policy, we use a SASS microbenchmark (see Listing~\ref{lst:tree_plru}). The benchmark first fills all four ways of a single set with unique addresses ($A, B, C$, and $D$). It then accesses these lines in different orders (the touch pattern) before inserting a fifth address ($E$) to force an eviction. By measuring which address survives, we infer the underlying state machine. If the hardware employed a standard LRU policy, touching $A$ in Phase 2 would leave $B$ as the oldest resident line, causing $B$ to be evicted when $E$ is allocated. However, our timing results show that touching $A$ consistently causes $C$ to be evicted. Through testing various access sequences, we determined that both the L0 and L1 constant caches implement an identical Tree-based Pseudo-LRU (Tree-PLRU) policy. 

The observed behavior is consistent with a 3-bit binary-tree state machine for a 4-way set. One root bit selects between two-way groups, and the corresponding child bit selects a victim within the chosen group. On each access, the corresponding bits flip to point away from the accessed path, thereby determining the pseudo-LRU eviction target for the next allocation.

\subsection{The GPC-Level Unified Cache (GCC)}
\label{subsec:gcc_cache}
When a constant load misses the SM-level caches, we observe a 57-cycle latency. This indicates a hit in the GPC Constant Cache (GCC), a structure that unifies constant data and instructions~\cite{ncu_profiling_guide}. Cross-SM microbenchmarks verify that an address fetched by one SM results in a hit in the GCC for any other SM residing within the same Graphics Processing Cluster (GPC). This confirms that the GCC is physically located within the GPC and is shared across all its constituent SMs. We therefore classify it as the L1.5 Cache.

\textbf{Extracting GCC Characteristics via Instruction Insertion.}
We first probe address offsets after an L1 eviction, revealing a 256\,B GCC line size. Set-conflict measurements then establish the indexing function as $\text{set\_index} = (\text{address} / 256) \pmod{64}$, yielding 64 sets.

However, measuring the associativity and total capacity is challenging. Because the CUDA programming model limits user-accessible constant memory to 64\,KiB, it is impossible to overflow the GCC using only constant data, so MT4G cannot measure its capacity on H100~\cite{MT4G}. To solve this limitation, we follow prior work~\cite{huerta_ref_97,huerta_ref_47,huerta_ref_46} and leverage the unified nature of the GCC by using instructions to force data evictions. Specifically, a SASS microbenchmark first loads $c$ constant lines that map to one GCC set and then executes $k$ instruction lines that map to the same set, using a chain of branches with a 16\,KiB stride. A timed probe then checks whether each constant line is still in the GCC.

With up to eight lines in the set, no constant line is evicted, even when all eight are constant lines. With nine lines, for example eight constant lines and one instruction line, one constant line is evicted in every run. Because this bound does not depend on which line the replacement policy evicts, the GCC is 8-way set-associative, with a capacity of \(64 \times 8 \times 256\,\mathrm{B} = 128\,\mathrm{KiB}\), much larger than the SM-level constant caches.

\textbf{Non-Standard Replacement Policy.}
In mixed instruction and data microbenchmarks, the GCC replacement behavior deviates from standard LRU, FIFO, or Tree-PLRU policies: after eight constant lines fill a set, the next insertion does not evict the first filled line, and re-accessing any line other than this victim does not change the victim. We leave a detailed characterization of the GCC replacement policy to future work.

\subsection{{Reverse Engineering Other GPUs}}
\label{subsec:other_gpus_re}
{Sections~\ref{sec:l1_microarchitecture}--\ref{sec:constant_hierarchy} recover the memory hierarchy of H20. To show that \projname{} is hardware-agnostic, we apply the same microbenchmarks to H100 (Hopper), A800 (Ampere), and RTX 4090 (Ada). As Table~\ref{tab:other_gpus_re} shows, all four GPUs share the organization we recover on H20 at every cache level and differ only in parameters, such as the L1 associativity and the number of L2 slices. The only structural difference is that RTX 4090 has a single L2 partition. Its hash still folds eight logical FBPs onto six physical FBPs, but the folding target rotates every 32\,KiB instead of every 128\,KiB.}

{
\begin{table}[t]
\centering%
\caption{{Reverse-engineered memory-hierarchy parameters on four GPUs.}}
\label{tab:other_gpus_re}
\footnotesize
\setlength{\tabcolsep}{3.4pt}
\renewcommand{\arraystretch}{1.05}
\begin{tabular}{@{}llcccc@{}}
\toprule
 & \textbf{Parameter} & \textbf{H20} & \textbf{H100} & \textbf{A800} & \textbf{RTX 4090} \\
\midrule
\textbf{L1D} & Slices$\times$sets$\times$ways & 4$\times$128$\times$4 & 4$\times$128$\times$4 & 4$\times$64$\times$6 & 4$\times$64$\times$4 \\
 & L1D+SMEM (KiB) & 249/256 & 249/256 & 185/192 & 121/128 \\
 & Hit latency (cyc.) & 31 & 31 & 31 & 32 \\
 & Index, replacement & \multicolumn{4}{c}{XOR, SLRU} \\
\midrule
\textbf{L2} & Part.$\times$FBPs$\times$slices & 2$\times$6$\times$8 & 2$\times$5$\times$8 & 2$\times$5$\times$4 & 1$\times$6$\times$6 \\
 & GPCs per partition & 4, 3 & 4, 4 & 4, 3 & \NAcell \\
 & FBP remap & 8$\to$6 & 8$\to$5 & 8$\to$5 & 8$\to$6 \\
 & Slice remap & 128$\to$96 & 128$\to$80 & 64$\to$40 & 64$\to$36 \\
 & Ways, replacement & \multicolumn{4}{c}{16, LRU} \\
 & Coherence & \multicolumn{3}{c}{Set-level} & \NAcell \\
\midrule
\textbf{Const.} & L0C / L1C & \multicolumn{4}{c}{2\,KiB, 8$\times$4, Tree-PLRU} \\
 & GCC sets$\times$ways & \multicolumn{4}{c}{64$\times$8} \\
\bottomrule
\end{tabular}
\end{table}
}

%% file: sections/modeling.tex
\section{Modeling and Evaluation}
\label{sec:modeling_eval}

{To evaluate the architectural impact of our reverse-engineered memory hierarchy, we integrate the recovered mechanisms into AccelSim and validate the simulation results against real hardware. Specifically, we use AccelSim v1.3.0~\cite{accel_sim} as extended by the open-source release of Huerta et al.~\cite{huerta2025}.}

\subsection{Supporting Real-World LLM Benchmarks}

Prior cycle-level GPU simulators have been evaluated primarily on traditional benchmark suites and CUTLASS GEMM workloads~\cite{gpgpu_sim,accel_sim,huerta2025}. However, modern LLM workloads extend far beyond GEMM, and performance simulation of their diverse operators has received little attention. To broaden this coverage, we construct \projname{}-Bench with 25 traditional GPU workloads widely used in prior work~\cite{gpgpu_sim,accel_sim,huerta2025} and 28 LLM operators. These operators span attention~\cite{vaswani2017attention}, normalization and activation~\cite{ba2016layer,zhang2019root,hendrycks2016gaussian,shazeer2020glu}, KV-cache management, token processing, quantization, and MoE. We take their implementations from vLLM~\cite{kwon2023efficient}, PyTorch~\cite{paszke2019pytorch}, and FlashAttention-3~\cite{shah2024flashattention}, with tensor dimensions configured based on the Qwen3-30B-A3B model~\cite{qwen3_30b_a3b}. The complete benchmark suite is listed in Table~\ref{tab:detailed_errors}.

\subsection{{Evaluation Methodology}}
\label{subsec:modeling_recovered_hierarchy}
{\textbf{Hardware measurement.} We measure the execution cycles of each workload with Nsight Compute. Nsight Compute locks the GPU clocks and flushes all caches before each kernel, and the simulator likewise starts each kernel with cold caches. We repeat each measurement five times and use the average.}

{\textbf{Baselines.} The simulator released with~\cite{huerta2025} does not cover the GPUs we evaluate. We therefore define a baseline (BL) with their RTX A6000 model, the closest of these models to our GPUs, and set its SM count, clock frequency, and shared-memory size to those of each target GPU (e.g., 78 SMs for H20). All other parameters keep their RTX A6000 values, such as the 6\,MiB L2 cache (Table~\ref{tab:sim_config}). We then improve it (denoted as BL-Impv) by further aligning its memory parameters with those documented by NVIDIA for each GPU, such as the 60\,MiB L2 cache and the 4\,TB/s memory bandwidth of H20. BL-Impv includes none of our reverse-engineered mechanisms, so any accuracy gain over BL-Impv comes from our findings.}

{\textbf{Configurations.} To isolate the contribution of each recovered cache level, we incrementally extend BL-Impv with three configurations. The L1D configuration replaces the L1 data cache of BL-Impv with our recovered model, L1D+L2D further replaces the L2, and \projname{} further adds the recovered constant-cache hierarchy, so it includes all the mechanisms we recover. Table~\ref{tab:sim_config} lists the parameters of each configuration.}

{
\begin{table}[t]
\centering%
\caption{{Memory-hierarchy parameters of the simulated configurations for H20.}}
\label{tab:sim_config}
\footnotesize
\setlength{\tabcolsep}{3.5pt}
\renewcommand{\arraystretch}{1.0}
\begin{tabular}{@{}llccc@{}}
\toprule
 & \textbf{Parameter} & \textbf{BL} & \textbf{BL-Impv} & \textbf{\projname{}} \\
\midrule
\textbf{L1D} & Slices$\times$sets$\times$ways & 1$\times$4$\times$256 & 1$\times$4$\times$512 & 4$\times$128$\times$4 \\
 & Capacity & 128\,KiB & 256\,KiB & 249\,KiB$-$SMEM \\
 & Index, replacement & Linear, LRU & Linear, LRU & XOR, SLRU \\
\midrule
\textbf{L2} & Slices & 48 & 48 & 96 \\
 & Partitions & 1 & 1 & 2, GPC affinity \\
 & Capacity & 6\,MiB & 60\,MiB & 60\,MiB \\
 & Slice hash & IPOLY & IPOLY & Sec.~\ref{subsec:l2_hash_recovery} \\
 & DRAM bandwidth & 2\,TB/s & 4\,TB/s & 4\,TB/s \\
\midrule
\textbf{Const.} & L0C / L1C size & 0.5 / 2\,KiB & 0.5 / 2\,KiB & 2 / 2\,KiB \\
 & L0C / L1C policy & LRU & LRU & Tree-PLRU \\
 & GCC & \NAcell & \NAcell & 64$\times$8 per GPC \\
\bottomrule
\end{tabular}
\end{table}
}

\subsection{Performance Accuracy}
\label{subsec:perf_accuracy}

{We simulate all 53 workloads under BL, BL-Impv, and the three configurations L1D, L1D+L2D, and \projname{}. Table~\ref{tab:detailed_errors} reports hardware cycles and per-workload cycle errors. Overall MAPE is 36.83\% with BL. Applying the documented memory parameters reduces it to 27.58\% with BL-Impv, and our recovered models further reduce it to 27.36\% with L1D, 20.62\% with L1D+L2D, and 20.04\% with \projname{}. Each incremental model improves aggregate accuracy, with distributed L2 modeling providing the largest reduction.}

\begin{figure}[t]
    \centering
    \includegraphics[width=\linewidth]{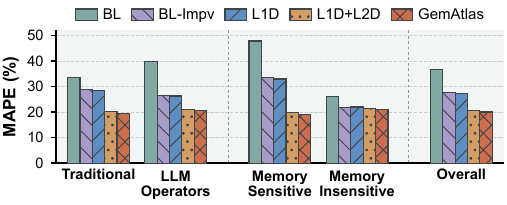}
    \caption{{Mean absolute percentage error (MAPE) on H20 workload groups.}}
    \label{fig:evaluation_mape}
\end{figure}

{Compared with BL-Impv, \projname{} reduces MAPE for both traditional benchmarks, from 28.69\% to 19.31\%, and LLM operators, from 26.59\% to 20.69\% (Figure~\ref{fig:evaluation_mape}).}
{To relate accuracy gains to memory sensitivity, we adapt Alsop et al.'s workload classification~\cite{alsop2019gpu_cache}: a workload is memory-sensitive if its simulated cycles change by at least 5\% between adjacent configurations from BL-Impv to \projname{}. This classification identifies 26 sensitive workloads, marked with an asterisk in Table~\ref{tab:detailed_errors}, and 27 insensitive workloads. As Figure~\ref{fig:evaluation_mape} shows, compared with BL-Impv, \projname{} reduces MAPE from 33.46\% to 19.12\% for sensitive workloads, but only from 21.91\% to 20.91\% for insensitive workloads. This contrast indicates that the simplified memory hierarchy in BL-Impv is a major source of error for sensitive workloads, while the recovered model corrects much of this error. The 20.91\% MAPE that remains for insensitive workloads, whose cycles change little across configurations, therefore comes mainly from parts of the simulator other than the memory hierarchy.}

\begin{figure}[t]
    \centering
    \includegraphics[width=0.9\columnwidth]{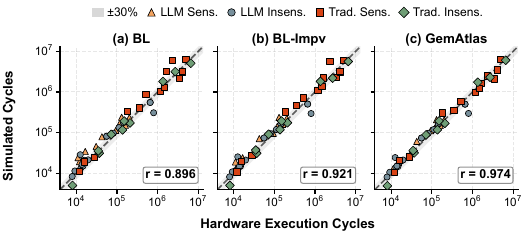}
    \caption{{Pearson correlation between simulated and hardware execution cycles on H20 under BL, BL-Impv, and \projname{}.}}
    \label{fig:evaluation_correlation}
\end{figure}

{Finally, \projname{} raises the Pearson correlation with hardware from 0.896 with BL and 0.921 with BL-Impv to 0.974 (Figure~\ref{fig:evaluation_correlation}), and increases the number of workloads within the $\pm 30\%$ error bound from 34 and 37 to 42, demonstrating stronger global agreement with hardware.}

{\textbf{Other GPUs.} We repeat the evaluation on H100, A800, and RTX 4090, building each configuration from the documented memory parameters of the GPU and our measurements in Sections~\ref{sec:l1_microarchitecture}--\ref{sec:constant_hierarchy}. As Table~\ref{tab:other_gpus} shows, \projname{} is more accurate than BL-Impv on every GPU, reducing MAPE from 37.7\% to 21.3\% on H100, from 32.5\% to 28.0\% on A800, and from 29.0\% to 19.7\% on RTX 4090. On average, \projname{} reduces MAPE from 34.04\% to 23.98\% for the LLM operators and from 36.33\% to 20.61\% for the memory-sensitive workloads.} %

{
\begin{table}[t]
\centering%
\caption{{MAPE (\%) on four GPUs.%
}}
\label{tab:other_gpus}
\footnotesize
\setlength{\tabcolsep}{5pt}
\renewcommand{\arraystretch}{1.05}
\begin{tabular}{@{}ccccc@{}}
\toprule
\textbf{GPU} & \textbf{BL-Impv} & \textbf{L1D} & \textbf{L1D+L2D} & \textbf{\projname{}} \\
\midrule
H20 & 27.6 & 27.4 & 20.6 & 20.0 \\
H100 & 37.7 & 37.4 & 21.9 & 21.3 \\
A800 & 32.5 & 32.3 & 29.3 & 28.0 \\
RTX 4090 & 29.0 & 28.5 & 20.2 & 19.7 \\
\midrule
Average & 31.7 & 31.4 & 23.0 & 22.2 \\
\bottomrule
\end{tabular}
\end{table}
}

\newcommand{\ErrGroup}[1]{\rowcolor{black!8}[0pt][0pt]\multicolumn{7}{@{}l@{}}{\textbf{#1}}\\}
\begin{table}[!t]
\centering
\caption{Per-workload H20 hardware cycles and simulation errors (\%). Asterisks mark memory-sensitive workloads.}
\label{tab:detailed_errors}
\scriptsize
\setlength{\tabcolsep}{4pt}
\renewcommand{\arraystretch}{0.95}
\resizebox{\columnwidth}{!}{%
\begin{tabular}{@{}lrrrrrr@{}}
\toprule
\textbf{Workload} & \textbf{HW cyc.} & \textbf{BL} & \textbf{{BL-Impv}} & \textbf{{L1D}} & \textbf{L1D+L2D} & \textbf{\projname{}} \\
\midrule
\ErrGroup{Traditional Benchmarks}
rodinia-pathfinder & 36726 & -14.98 & {-15.21} & {-15.23} & -14.50 & -13.67 \\
rodinia-hotspot & 72095 & +25.63 & {+26.81} & {+27.06} & +23.28 & +22.05 \\
rodinia-bfs & 137208 & -15.07 & {-13.99} & {-14.03} & -14.30 & -14.40 \\
rodinia-nn\textsuperscript{*} & 27681 & -11.11 & {-0.92} & {-1.12} & -7.54 & -7.81 \\
rodinia-srad-v2 & 34740 & +2.07 & {+7.38} & {+7.96} & +4.72 & +4.28 \\
rodinia-streamcluster\textsuperscript{*} & 1162650 & -10.12 & {-5.43} & {-5.77} & -14.63 & -14.85 \\
rodinia-backprop\textsuperscript{*} & 15598 & +23.44 & {+27.54} & {+26.84} & +34.41 & +33.85 \\
rodinia-lud & 209332 & -17.48 & {-16.28} & {-16.31} & -19.19 & -19.01 \\
rodinia-heartwall\textsuperscript{*} & 11871 & -7.35 & {-7.90} & {-7.49} & -4.51 & -9.82 \\
rodinia-nw & 150396 & +28.41 & {+28.49} & {+28.49} & +28.44 & +28.53 \\
parboil-bfs & 8056 & -37.70 & {-37.60} & {-37.60} & -39.09 & -38.93 \\
parboil-mri-q\textsuperscript{*} & 349190 & +17.44 & {+31.99} & {+32.11} & +18.70 & +2.24 \\
parboil-histo\textsuperscript{*} & 3403026 & -35.05 & {-33.21} & {-33.08} & -38.55 & -38.50 \\
parboil-stencil & 6551610 & -22.47 & {-14.30} & {-14.30} & -10.87 & -8.01 \\
parboil-spmv\textsuperscript{*} & 1619775 & +96.04 & {+37.98} & {+37.54} & -0.99 & -1.24 \\
parboil-lbm\textsuperscript{*} & 2322409 & +162.46 & {+145.20} & {+145.81} & +7.08 & +7.05 \\
polybench-atax\textsuperscript{*} & 4114959 & -23.51 & {-23.37} & {-22.88} & -15.52 & -15.76 \\
polybench-bicg\textsuperscript{*} & 4116584 & -23.40 & {-23.67} & {-22.58} & -15.59 & -16.00 \\
polybench-mvt\textsuperscript{*} & 4093132 & -22.68 & {-22.94} & {-22.06} & -15.13 & -14.71 \\
polybench-2dconv\textsuperscript{*} & 182170 & +81.69 & {+77.54} & {+78.07} & +67.57 & +67.93 \\
polybench-3dconv\textsuperscript{*} & 1465403 & -10.98 & {-7.57} & {-7.63} & -14.87 & -14.79 \\
polybench-gemm\textsuperscript{*} & 515826 & +68.74 & {+38.89} & {+38.77} & +25.69 & +16.09 \\
polybench-gesummv & 2708664 & +16.58 & {-19.25} & {-15.47} & -14.99 & -15.05 \\
lonestar-bfs-wla-rmat20\textsuperscript{*} & 4930879 & +26.55 & {+19.77} & {+21.45} & +23.62 & +30.10 \\
pannotia-fw\_block-512 & 1356172 & +33.77 & {+33.92} & {+34.21} & +28.11 & +27.98 \\
\midrule
\textbf{Traditional MAPE} & -- & \textbf{33.39} & \textbf{{28.69}} & \textbf{{28.55}} & \textbf{20.08} & \textbf{19.31} \\
\midrule
\ErrGroup{LLM Operators}
vllm-rmsnorm & 12556 & +128.27 & {+101.91} & {+103.21} & +97.69 & +93.95 \\
vllm-fused-add-rmsnorm\textsuperscript{*} & 16583 & +106.81 & {+45.90} & {+44.82} & +26.26 & +24.91 \\
vllm-layernorm\textsuperscript{*} & 118510 & +46.37 & {+21.23} & {+21.57} & +9.79 & +9.37 \\
torch-layernorm & 19537 & +9.50 & {+8.20} & {+9.03} & +8.65 & +8.44 \\
vllm-silu-mul & 81949 & +18.20 & {+14.93} & {+15.60} & +16.84 & +17.23 \\
torch-swiglu & 111998 & +18.77 & {+6.48} & {+6.39} & +6.12 & +6.44 \\
torch-ffn-gate & 798534 & -61.46 & {-63.08} & {-63.07} & -62.97 & -62.91 \\
torch-embedding\textsuperscript{*} & 12017 & +61.48 & {+23.67} & {+22.90} & +0.88 & -2.86 \\
torch-softmax-vocab\textsuperscript{*} & 93853 & +20.85 & {+2.39} & {+5.68} & +19.43 & +19.24 \\
torch-topk-vocab\textsuperscript{*} & 201720 & +16.32 & {+11.28} & {+11.06} & +1.36 & +2.12 \\
torch-gelu\textsuperscript{*} & 30833 & +48.12 & {+10.88} & {+10.69} & +1.72 & +1.68 \\
torch-rope\textsuperscript{*} & 161187 & -6.06 & {-0.70} & {-0.51} & -6.78 & -6.92 \\
vllm-paged-attn-v2\textsuperscript{*} & 130354 & +83.63 & {+81.43} & {+73.22} & +60.40 & +59.78 \\
flash3-mqa-kvcache & 55781 & +45.35 & {+31.71} & {+31.71} & +33.28 & +31.90 \\
flash3-packgqa-kvcache & 55699 & +46.04 & {+32.63} & {+32.73} & +32.54 & +32.49 \\
vllm-reshape-cache & 9224 & +23.69 & {+17.61} & {+16.90} & +15.80 & +14.60 \\
vllm-reshape-cache-flash & 7915 & +5.19 & {+3.10} & {+3.01} & +1.97 & +2.21 \\
vllm-token-quant-fp8\textsuperscript{*} & 15169 & +28.12 & {+29.72} & {+22.01} & +28.47 & +21.76 \\
vllm-token-group-quant & 15685 & +33.04 & {+15.04} & {+15.07} & +15.39 & +12.98 \\
vllm-silu-mul-fp8 & 145461 & +22.28 & {+11.73} & {+11.16} & +11.90 & +12.68 \\
vllm-2d-silu-mul-fp8 & 144148 & +23.35 & {+12.87} & {+13.41} & +13.37 & +13.30 \\
vllm-fused-topk & 13052 & +4.68 & {+4.60} & {+5.14} & +4.51 & +6.31 \\
torch-residual-add\textsuperscript{*} & 11038 & +120.19 & {+72.01} & {+73.87} & +17.64 & +15.24 \\
vllm-moe-align-block & 14670 & +2.95 & {+0.95} & {-0.32} & -1.21 & -0.26 \\
vllm-moe-permute\textsuperscript{*} & 42660 & +85.31 & {+66.95} & {+67.01} & +40.91 & +42.60 \\
vllm-moe-unpermute & 13834 & +19.38 & {+6.68} & {+9.66} & +6.90 & +9.38 \\
vllm-grouped-gemm & 655735 & -15.83 & {-23.13} & {-23.13} & -24.04 & -23.81 \\
vllm-cutlass-moe-fp8 & 658249 & -16.21 & {-23.60} & {-23.49} & -23.98 & -23.90 \\
\midrule
\textbf{LLM MAPE} & -- & \textbf{39.91} & \textbf{{26.59}} & \textbf{{26.30}} & \textbf{21.10} & \textbf{20.69} \\
\midrule
\textbf{Overall MAPE} & -- & \textbf{36.83} & \textbf{{27.58}} & \textbf{{27.36}} & \textbf{20.62} & \textbf{20.04} \\
\bottomrule
\end{tabular}%
}
\end{table}

\subsection{Impact of L1D Modeling}
\label{subsec:l1d_modeling_impact}
{Replacing the L1D of BL-Impv with the recovered H20 L1D changes overall MAPE only slightly, from 27.58\% to 27.36\%. Hardware L1 hit rates measured with Nsight Compute explain why: 22 of the 53 workloads hit in the L1 for fewer than 10\% of their global loads, and for 21 others both BL-Impv and L1D already match the hardware hit rate within 2 percentage points. In 20 of these 21 workloads, more than 99\% of the re-accessed sectors in the traces are re-accessed after fewer than 64\,KiB of other data, a quarter of the L1 capacity, so both L1 organizations retain them. Only \texttt{polybench-gesummv} depends on the L1 organization. Its warps access memory with a 16\,KiB stride and re-access data after 64 to 128\,KiB. The linear index of BL-Impv maps these accesses to the same set and causes set conflicts, whereas the recovered XOR index reduces these conflicts, so its hit rate rises from 86.10\% to 90.22\%. Its cycle error also drops from $-19.25\%$ to $-15.47\%$.}

\subsection{Impact of Distributed L2 Modeling}
\label{subsec:l2d_modeling_impact}

{Distributed L2D modeling provides the largest incremental accuracy improvement, reducing overall MAPE from 27.36\% with L1D to 20.62\% with L1D+L2D. Starting from the L1D configuration, we add three components of the recovered L2 model step by step. (1) The slice count replaces the 48 slices of BL-Impv with the 96 slices recovered in Section~\ref{subsec:l2_routing}. (2) The FBP/slice routing hash replaces the IPOLY hash and selects the partition, FBP, and slice of each address. (3) The partition model adds the two partitions, the GPC-to-partition mapping, the measured local and cross-partition latencies, and cross-partition coherence, which together form the partitioned L2 of H20 (Sections~\ref{subsec:l2_partition} and~\ref{subsec:l2_es_coherence}). We do not split this component further because a model with only some of these mechanisms does not correspond to real hardware. The slice count reduces MAPE to 22.14\%, the largest reduction among the three components. With only 48 slices, the simulated L2 serves fewer requests in parallel, so L1D overestimates execution cycles by 14.30\% on average, compared with 6.87\% with 96 slices. Adding routing hash then lowers MAPE slightly to 21.83\%, and the partition model finally reduces the MAPE to 20.62\% by accounting for cross-partition accesses. In \texttt{parboil-stencil}, for example, 46\% of the L2 reads come from SMs in the other partition, and modeling their latency and replication reduces the error from $-29.3\%$ to $-10.9\%$.}

\subsection{Impact of Constant-Memory Modeling}
\label{subsec:constant_modeling_impact}

Adding the recovered constant-memory hierarchy reduces overall MAPE from 20.62\% to 20.04\%. {The gain is small because constant-memory accesses exceed 1\% of L2 accesses in only 9 of the 53 workloads. Nevertheless, \projname{} reduces the constant-memory requests reaching the L2 in 52 of the 53 workloads, by 71.15\% in total. In \texttt{parboil-mri-q}, which has the highest constant-memory share (6.16\%), the GCC reduces these requests by 95.35\% and the cycle error from $+18.70\%$ to $+2.24\%$.}

%% file: sections/discussion.tex
{
\section{Discussion}
\label{sec:discussion}

Beyond improving simulation accuracy, the mechanisms we recover provide architectural insights for both software optimization and hardware design.

\textbf{L2-aware data placement.} Eq.~(\ref{eq:l2_hash_full}) gives the partition of every address, so software can place the data that each SM reads in the SM's local partition. This avoids both the extra latency of cross-partition accesses (Section~\ref{subsec:l2_partition}) and the L2 capacity taken by their coherent copies (Section~\ref{subsec:l2_es_coherence}). We apply this placement to the vLLM paged-attention decode kernel on H20. It removes 99\% of the cross-partition requests, lets more KV data fit in the L2, and speeds up the kernel by 1.10--1.12$\times$ over the same kernel without this placement. The hash also suggests placement at the FBP level: because of the rotating remap in Eq.~(\ref{eq:remap}), the FBP load is balanced only over three consecutive 128\,KiB ranges, so a hot region of about 128\,KiB could be spread over 384\,KiB to use all FBPs evenly.

\textbf{Decoupled L1 tags and data.} The recovered L1 decouples its tag and data arrays, which appears to let it switch capacity between the L1 and SMEM. The tag array is fixed, so the L1 keeps the same XOR index under every SMEM configuration, while the data array shrinks as the SMEM allocation grows (Section~\ref{subsec:4slice_and_asymmetry}). The data array can therefore fill up before the tags do, and the L1 needs cross-set eviction to make room, similar to the global data replacement of the V-Way cache~\cite{qureshi2005vway}. Cross-set eviction starts when the L1 data and SMEM together hold 249\,KiB, 7\,KiB below the 256\,KiB unified capacity (Table~\ref{tab:l1_capacity}). We conjecture that this 7\,KiB is a buffer that holds incoming lines when cross-set eviction cannot keep up with heavy fill traffic. Because data entries are allocated independently of the tags, we suggest a possible design that may improve performance: one tag owns several 128\,B data entries as a larger line, and all of these entries are released when the tag is evicted. Data with spatial locality could then use larger lines while other data keeps 128\,B lines, without adding tags. We leave the evaluation of this design to future work.
\par}

%% file: sections/relatedwork.tex
\section{Related Work}
\label{sec:relatedwork}

Prior work has studied GPU caches and TLBs from GT200 to A100~\cite{huerta_ref_97,mei2017dissecting,li2014swcache,huerta_ref_52,huerta_ref_47,huerta_ref_46,huerta_ref_48,huerta_ref_104} and GPU NoCs~\cite{huerta_ref_4,noc_paper}, and, more recently, Hopper-era features, automated topology discovery, and the L2/DRAM NUMA topology~\cite{huerta_ref_58,MT4G,liu2026dgna}. These studies provide component-level visibility but do not reconstruct a unified, cache-set-level hierarchy. {We next compare our methods and architectural findings with these studies.}

{Our probes build on group elimination~\cite{cite:group_elimination}, bit-flip analysis~\cite{bitflip}, and GCC eviction with instruction fetches~\cite{huerta_ref_97,huerta_ref_47,huerta_ref_46}. Our SASS probes measure latency more accurately than the CUDA/PTX benchmarks of prior work~\cite{huerta_ref_58,noc_paper,MT4G,liu2026dgna}: our probes time only the load, whereas these benchmarks also time cold misses or dependent instructions. We also introduce five new methods: (1) SASS-level eviction sets for the L1 and constant caches, (2) L1 capacity measurement at cache-line granularity instead of CUDA pointer chasing~\cite{mei2017dissecting,MT4G}, (3) reading the SM-to-GPC and GPC-to-partition mappings from special registers instead of inferring them from latencies and thread-block clusters~\cite{huerta_ref_48,noc_paper,liu2026dgna}, (4) grouping L2 slices into FBPs by latency correlation, and (5) slice-constrained group elimination, to our knowledge the first method to reliably and efficiently construct eviction sets on partitioned GPU L2 caches, where prior techniques fail~\cite{cite:group_elimination,invalidate}.

Several of our architectural findings refine or extend prior measurements, and Table~\ref{tab:wide_table} marks them as improved (I). For the L1, we improve three prior results. First, Jia et al. find the L1 capacity of Volta and Turing 7\,KiB below the nominal size and conjecture that a new replacement policy causes this gap~\cite{huerta_ref_47,huerta_ref_46}, whereas our line-by-line probe finds the same gap on H20 before any set holds more than four lines, which rules out replacement within a set (Section~\ref{sec:discussion}). Second, prior work reports the L1 associativity only for Fermi~\cite{mei2017dissecting}, and to our knowledge we are the first to measure it on a modern GPU (4-way on H20). Third, the benchmarks of Luo et al. and MT4G yield L1 hit latencies of 41 and 38 cycles on H20~\cite{huerta_ref_58,MT4G}, whereas our SASS probe measures a more accurate 31 cycles. For the L2, we improve two prior results. First, Jia et al. report a 16-way L2 on Volta and Turing~\cite{huerta_ref_47,huerta_ref_46}, and to our knowledge we are the first to measure the associativity of a partitioned L2 (16-way on H20). Second, Mei and Chu conclude from aperiodic miss patterns that the Fermi L2 does not use LRU, and note that its address mapping is irregular~\cite{mei2017dissecting}. Our eviction sets access a single set, which separates the replacement policy from the address mapping, and show that H20 uses LRU. For the constant caches, we improve three prior results. First, prior work reports a 2\,KiB, 4-way L1 constant cache with 64\,B lines and 8 sets on GT200, Volta, and Turing~\cite{huerta_ref_97,huerta_ref_47,huerta_ref_46}, but only its capacity and line size on H100~\cite{MT4G}, whereas we find the same geometry on H20 together with its linear set index. Second, prior work measures a single hit latency per constant-cache level~\cite{huerta_ref_97,huerta_ref_47,huerta_ref_46,huerta_ref_48,MT4G}, whereas we find that the latency also depends on the instruction: \texttt{MOV} and \texttt{ULDC} both hit in the L0 but take 2 and 15 cycles. Third, Jia et al. measure the GCC capacity on Volta, Turing, and A100~\cite{huerta_ref_47,huerta_ref_46,huerta_ref_48}, and MT4G cannot measure it beyond 64\,KiB on H100 because CUDA limits constant memory to 64\,KiB~\cite{MT4G}. Evicting GCC lines with instruction fetches~\cite{huerta_ref_97,huerta_ref_47,huerta_ref_46}, we fill a single set with both constant and instruction lines and find 64 sets of 8 ways (128\,KiB).

The remaining findings, marked as new (N) in Table~\ref{tab:wide_table}, have not been reported before to our knowledge. For the L1, they are the four slices, the XOR set index, the tag--data asymmetry that causes cross-set evictions, and the SLRU policy. These findings also explain two observations of Jia et al. on Volta and Turing~\cite{huerta_ref_47,huerta_ref_46}: the four L1 sets they report appear to be the four slices, and the random evictions they observe appear to be cross-set evictions. For the L2, they are the partition/FBP/slice hash with its 128-to-96 remap and the set-level coherence, whereas prior work observes the address interleaving between the two partitions and characterizes coherence at cache-line granularity~\cite{huerta_ref_48,liu2026dgna}. For the constant caches, they are the L0 shared by \texttt{MOV} and \texttt{ULDC} with its geometry and index, the GCC index, and the Tree-PLRU policy of the L0 and L1, whereas prior work identifies the L0~\cite{huerta2025} and finds only that the L1 does not use LRU~\cite{huerta_ref_47,huerta_ref_46}.\par}
These memory-system gaps also affect GPU simulators such as GPGPU-Sim~\cite{gpgpu_sim}, Accel-Sim~\cite{accel_sim,huerta_ref_52}, gem5 and its GPU models~\cite{huerta_ref_20,huerta_ref_45,huerta_ref_81}, and MGPUSim~\cite{huerta_ref_91}, and recent work revisits their hardware assumptions~\cite{huerta_ref_43,huerta2025}. However, existing simulators lack hardware-faithful slice routing, set indexing, replacement, and cross-partition coherence.

%% file: sections/conclusion.tex
\section{Conclusion}
In this paper, we present \projname{}, a systematic reverse engineering study of the memory hierarchy of modern NVIDIA GPUs. We develop a SASS-level microbenchmarking suite to recover undocumented microarchitectural behavior. We characterize the L1 data cache in depth (including sliced organization, XOR-based routing, tag/data asymmetry, and segmented replacement), reconstruct the L2 partition/slice/FBP hierarchy, recover address hash and remap logic, expose cache-set granularity cross-partition coherence and replacement behavior, and introduce an efficient hash-guided slice-constrained, eviction-set construction method for modern GPUs. We further characterize the constant-memory hierarchy, covering its fine-grained geometry and replacement policy. 
Finally, incorporating these findings into a cycle-level GPU simulator reduces the average MAPE {over four GPUs} from {34.04\%} to {23.98\%} for real-world LLM operators, from {36.33\%} to {20.61\%} for memory-sensitive workloads, and from {31.69\%} to {22.24\%} overall.